\documentclass{article}

\usepackage{arxiv}

\usepackage[utf8]{inputenc} 
\usepackage[T1]{fontenc}    
\usepackage{hyperref}       
\usepackage{url}            
\usepackage{booktabs}       
\usepackage{amsfonts}       
\usepackage{nicefrac}       
\usepackage{microtype}      
\usepackage{lipsum}
\usepackage{graphicx}
\usepackage{fancyhdr}
\usepackage[most]{tcolorbox}
\usepackage{pgfplots}
\usepackage{longtable}
\graphicspath{ {./images/} }

\usepackage{booktabs}      
\usepackage{tabularx}      
\usepackage{colortbl}      
\usepackage{xcolor}        
\usepackage{array}         

\definecolor{rowgray}{gray}{0.93}
\newcolumntype{L}[1]{>{\raggedright\arraybackslash}p{#1}}
\newcolumntype{Y}{>{\raggedright\arraybackslash}X}

\newcommand{\threat}{\sloppy \textsf{\textsc{\small{Threat}}}}
\newcommand{\threats}{\sloppy \textsf{\textsc{\small{Threats}}}}
\newcommand{\threatElement}{\sloppy \textsf{\textsc{\small{Threat Element}}}}
\newcommand{\threatElements}{\sloppy \textsf{\textsc{\small{Threat Elements}}}}
\newcommand{\threatActor}{\sloppy \textsf{\textsc{\small{Threat Actor}}}}
\newcommand{\countermeasure}{\sloppy \textsf{\textsc{\small{Countermeasure}}}}
\newcommand{\countermeasures}{\sloppy \textsf{\textsc{\small{Countermeasures}}}}
\newcommand{\countermeasureElement}{\sloppy \textsf{\textsc{\small{Countermeasure Element}}}}
\newcommand{\countermeasureElements}{\sloppy \textsf{\textsc{\small{Countermeasure Elements}}}}
\newcommand{\mitigates}{\sloppy\textsf{\textsc{\small{mitigates}}}}
\newcommand{\fulfills}{\sloppy \textsf{\textsc{\small{fulfills}}}}
\newcommand{\implements}{\sloppy \textsf{\textsc{\small{implements}}}}
\newcommand{\securityProperty}{\sloppy \textsf{\textsc{\small{Security Property}}}}
\newcommand{\securityProperties}{\sloppy \textsf{\textsc{\small{Security Properties}}}}
\newcommand{\stride}{\sloppy \textsf{\textsc{\small{STRIDE}}}}

\newcommand{\sr}{\sloppy \textsf{\textsc{\small{SR}}}}
\newcommand{\srs}{\sloppy \textsf{\textsc{\small{SRs}}}}

\newcommand{\ams}{\sloppy \textsf{\textsc{\small{AMs}}}}
\newcommand{\ame}{\sloppy \textsf{\textsc{\small{AME}}}}
\newcommand{\ames}{\sloppy \textsf{\textsc{\small{AMEs}}}}
\newcommand{\ad}{\sloppy \textsf{\textsc{\small{AD}}}}
\newcommand{\ads}{\sloppy \textsf{\textsc{\small{ADs}}}}
\newcommand{\sa}{\sloppy \textsf{\textsc{\small{SA}}}}

\newcommand{\sad}{\sloppy \textsf{\textsc{\small{SAD}}}}
\newcommand{\sads}{\sloppy \textsf{\textsc{\small{SADs}}}}

\newcommand{\sds}{\sloppy \textsf{\textsc{\small{SDS}}}}
\newcommand{\sdss}{\sloppy \textsf{\textsc{\small{SDSs}}}}

\newcommand{\srd}{\sloppy \textsf{\textsc{\small{SRD}}}}
\newcommand{\traceabilityChain}{\sloppy \textsf{\textsc{\small{Traceability Chain}}}}

\usepackage{listings}

\lstdefinelanguage{OCL}{
  morekeywords={
    context,inv,self,let,in,if,then,else,endif,
    and,or,not,implies,
    select,exists,forAll,includes,notEmpty
  },
  sensitive=true,
  morecomment=[l]{--},
  morestring=[b]'
}

\lstdefinestyle{ocl}{
  language=OCL,
  basicstyle=\ttfamily\scriptsize,
  keywordstyle=\bfseries,
  columns=fullflexible,
  keepspaces=true,
  breaklines=true,
  showstringspaces=false,
  aboveskip=3pt,
  belowskip=3pt
}

\newcommand{\stereotype}{\sloppy \texttt{{{Stereotype}}}}
\newcommand{\stereotypes}{\sloppy \texttt{{{Stereotypes}}}}
\newcommand{\taggedValue}{\sloppy \texttt{{{Tagged Value}}}}
\newcommand{\taggedValues}{\sloppy \texttt{{{Tagged Values}}}}
\newcommand{\constraint}{\sloppy \texttt{{{Constraint}}}}
\newcommand{\constraints}{\sloppy \texttt{{{Constraints}}}}
\newcommand{\enumeration}{\sloppy \texttt{{{Enumeration}}}}
\newcommand{\enumerations}{\sloppy \texttt{{{Enumerations}}}}
\newcommand{\metaclass}{\sloppy \texttt{{{Metaclass}}}}
\newcommand{\metaclasses}{\sloppy \texttt{{{Metaclasses}}}}
\newcommand{\package}{\sloppy \texttt{{{Package}}}}
\newcommand{\component}{\sloppy \texttt{{{Component}}}}
\newcommand{\class}{\sloppy \texttt{{{Class}}}}
\newcommand{\dependency}{\sloppy \texttt{{{Dependency}}}}
\newcommand{\abstraction}{\sloppy \texttt{{{Abstraction}}}}

\title{SA-UML: Modeling Integrated Security Architectures in UML}

\author{
  Leon-Pascal Vidur Gawdi \\
  Software Architecture (SWA) Research Group\\
  RWTH Aachen University\\
  Ahornstraße 55 \\
  Aachen, Germany \\
  \texttt{leon.gawdi@rwth-aachen.de}
  \And
  Alex R. Mattukat \\
  Research Group Software Construction\\
  RWTH Aachen University\\
  Ahornstraße 55 \\
  Aachen, Germany \\
  \texttt{mattukat@swc.rwth-aachen.de}\\
}

\begin{document}
\tiny
This is the \textbf{peer-reviewed, accepted version} of a paper, including minor changes to content that was anonymized for double-blind review. It will appear in the proceedings of the 
15th International Conference on Model-Based Software and Systems Engineering (\textbf{MODELSWARD 2027}). The final published version will be available from Science and Technology Publications (SCITEPRESS).
© 2027 SCITEPRESS. Personal use of this material is permitted. Permission from SCITEPRESS must be obtained for all other uses, in any current or future media, including reprinting or republishing this material for advertising or promotional purposes, creating new collective works, resale or redistribution to servers or lists, or reuse of any copyrighted component of this work in other works.
\normalsize
\maketitle
\thispagestyle{fancy}
\begin{abstract}
Maintaining a security architecture description (SAD) separately from the overall architecture description (AD) can introduce redundancies and synchronization problems. Existing security modeling approaches typically address specific security aspects, domains, or analysis tasks rather than integrating SADs into ADs by design. 
This paper presents the Security Architecture UML profile (SA-UML), which enables modeling integrated SADs within ADs.
SA-UML represents security concepts and their architectural relationships through UML stereotypes, tagged values, and constraints within the overall AD. We demonstrate it in a case study modeling structural and behavioral views of a fictional document portal.
An exploratory questionnaire study ($n = 27$) examines how participants interpret predefined component diagrams. Recorded page-level completion times were shorter in the SA-UML condition, and participants perceived the approach as useful, while the response-accuracy comparison remained inconclusive. However, the time difference cannot be attributed solely to the notation, since the diagrams differed in their explicit security content and representation. Overall, our results are promising, but a more comprehensive evaluation is needed.
\end{abstract}
\newtcolorbox{mainBox}[1]{%
		enhanced,
		colback=white!90!black,
		colframe=darkgray,
		colbacktitle=darkgray,
		coltitle=white,
		fonttitle=\bfseries,
		title=#1,
		rounded corners,
		boxrule=3pt,
		drop shadow=darkgray,
		width=\textwidth,
	}


\section{\uppercase{Introduction}}\label{sec:intro}

According to the recently updated OWASP\footnote{Open Worldwide Application Security Project} Top Ten list, the de facto standard for the most critical security risks in web applications, risks stemming from insecure design remain among the six most frequent categories \cite{OWASP.Top10}. Yet, designing secure software is both crucial and challenging \cite{Gressl.2019}. Moreover, due to the shortage of security experts \cite{Furnell.2020}, non-security experts are often responsible for security-related activities and decision-making \cite{Gutfleisch.2022,Naji.2025}. This calls for greater support for software engineers and architects in designing secure software systems.

Secure system design is a central approach grounded in the ``secure by design'' principle for developing secure software systems. It involves designing the architecture's structure and behavior to ensure a software system's security, such as access control and denial-of-service prevention. We refer to the artifact produced in this design activity as the ``security architecture description'' (\sad) of the software system's ``security architecture'' (\sa), in line with the common term of an ``architecture description'' (\ad) referring to the design artifacts of an architecture \cite{rozanski}. The software system's \sa\ comprises the set of physical and logical security-related parts of the (overall) system architecture \cite{NIST800-37}. Thus, the \sad\ can be seen as a view of the overall system architecture that represents the system's \sa\ \cite{rozanski}.

In previous work~\cite{Sabau2024} we showed that although the \sa{} is an integral part of the overall architecture, the \sad{} is often maintained separately from the overall \ad{}, and identified three related challenges:


\begin{enumerate}
    \item \textbf{Redundancy:} Since security is highly cross-cutting, maintaining the \sad\ separately duplicates content already in the overall \ad, impairing the \ad's effectiveness as a communication tool.
    \item \textbf{Model Coupling:} Separating the \sad\ from the overall architecture description creates coupling between the two.
    \item \textbf{Lack of experts:} Designing \sads\ requires security experts. These are rare, and this shortage is not expected to change soon. 
\end{enumerate}


\paragraph{Contribution:} To address these shortcomings, we present a novel architecture modeling approach in~\cite {Sabau2024}. However, we did not implement it. In this paper, we close this gap by implementing the ``Security Architecture UML'' (SA-UML) profile.  SA-UML is a novel Unified Modeling Language (UML) profile that allows for modeling \sads\ integrated into overall \ads. Implemented as a UML profile, it fully conforms to the UML metamodel. This paper presents the implementation, a demonstration through a case study, and an trend analysis evaluation with $n=27$ participants. The UML profile and all evaluation results are published in the artifact repository~\cite{zenodo}. This study is based on the following research questions:

\vspace{5pt}

\noindent \textbf{RQ1:} How can an integrated security architecture modeling approach be implemented through a UML profile?

\vspace{5pt}

\noindent \textbf{RQ2:} Does the UML profile support practitioners in understanding security architecture descriptions?

\paragraph{Outline:} The rest of the paper is structured as such: Section~\ref{sec:related} presents related work. The conceptual foundations are introduced in section~\ref{sec:foundations}. In section~\ref{sec:method} we briefly describe our research approach. Section~\ref{sec:scam-dev} presents the conceptual and technical implementation of SA-UML. It is demonstrated in section~\ref{sec:demonstration} and evaluated in section~\ref{sec:evaluationSetup}. The results and limitations are discussed in section~\ref{sec:discussion}. Section~\ref{sec:conclusion} concludes the paper.
\section{Related Work}\label{sec:related}
Our approach is closely related to UMLsec~\cite{Jurjens2001}. UMLsec focuses on formally specifying and verifying security properties in UML models. However, it does not primarily address integrating \sads\ into the overall \ad. Moreover, UMLsec is no longer maintained and has been superseded by \textit{CARiSMA}~\cite{Carisma,CARiSMARepository}. CARiSMA provides tool support for analyzing UML models with the UMLsec profile. Compared to UMLsec, it also supports privacy analysis.
However, the project is more focused on analysis than on a dedicated modeling approach for integrated \sads.

Another approach to integrating security into modeling is \textit{SecureUML}~\cite{SecureUML}. This profile adds a UML-based modeling language for role-based access control.
However, it has a comparatively narrow security focus on access control and authorization, and does not cover broader security concerns.

\textit{SysML-Sec} extends the Systems Modeling Language (SysML) to integrate security requirements, threats, and security mechanisms into the development of embedded systems~\cite{ApvrilleRoudier2013}. Its methodology combines hardware/software partitioning, software design, and formal security verification. The supporting modeling environment, TTool\footnote{\url{https://ttool.telecom-paris.fr}}, transforms security design models into formal specifications for analysis with ProVerif, an automated verifier for security protocols~\cite{ApvrilleRoudier2013}.
SA-UML differs in that it uses a standard UML profile to organize \sads{} through UML packages, components, and classes, with an intended audience that includes software architects without security expertise.
This is a design motivation, not an empirically established advantage over SysML-Sec.

ArchiMate's \textit{Enterprise Risk and Security Modeling (ERSM)} addresses risk and security concerns at the enterprise architecture level \cite{Oliveira.2024}. While the approach allows for the integration of security-related data into the overall \ad, it differs from ours, as its focus lies on enterprise architectures, while ours focuses on the software and system architecture level. Moreover, our approach uses UML rather than ArchiMate as the architecture modeling language.

Overall, to our knowledge, no UML approach exists that integrates \sads\ into the overall \ad\ at the software or system architecture level.
\section{\uppercase{Foundations}}\label{sec:foundations}
\begin{figure*}[htb]
\centering
    \includegraphics[width=0.9\textwidth]{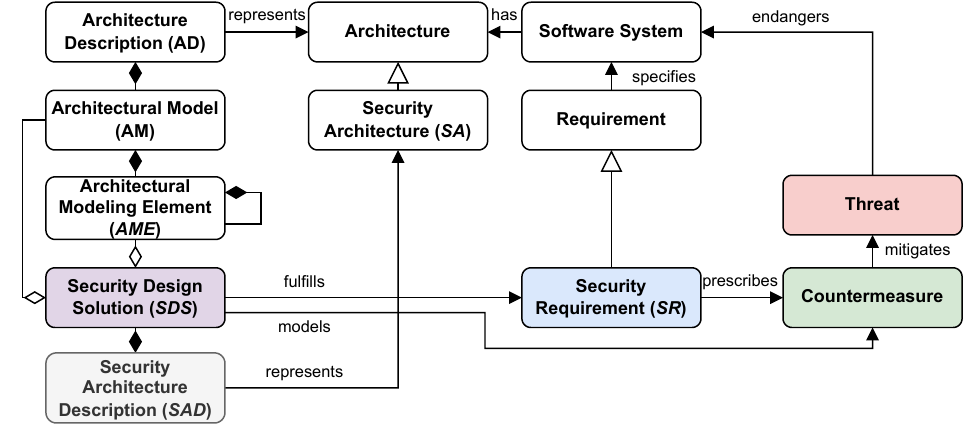}
    \caption{The Security Architecture Model (SAM), slightly modified from our proposal in \cite{Sabau2024}. Relationships follow UML class diagram notation. Colored nodes indicate recurring concepts in the SA-UML implementation.}
    \label{fig:concept-model}
\end{figure*}

SA-UML is based on a refinement of the concepts presented in~\cite{Sabau2024}. We have consolidated these into the SAM (Security Architecture Model) conceptual model. Figure \ref{fig:concept-model} depicts it. The refinements incorporate security concepts as defined by the National Institute of Standards and Technology (NIST) \cite{NIST_FIPS.200,NIST.800-63-4}.


We introduce \textsf{\textsc{\small{This Font}}} for central, conceptual terminology.  Moreover, we highlight five nodes in color to establish a color coding for use in figures. Both make central recurring terms in both text and images easier to recognize: central concepts are formatted in the text \textsf{\textsc{\small{As Such}}}, and use the same color throughout the figures. 

\paragraph{Security Concepts}

A \threat\ is a potential cause, such as a circumstance or event, that could endanger a software system. 

To reduce or fully mitigate \threats, different kinds of \countermeasures\ exist.  Examples are access control and denial-of-service prevention. 

\paragraph{Architectural Concepts}

Every software system has an architecture. A software system is specified by a set of requirements. A \textsf{\textsc{\small{security requirement (SR)}}} is a security-related requirement that prescribes the implementation of one or more \countermeasures.

The \ad\ documents an architecture. It is composed of a set of \textsf{\textsc{\small{architectural models (AMs)}}}, which are self-contained design documents, such as UML diagrams. Multiple atomic \textsf{\textsc{\small{architectural modeling elements (AMEs)}}}, such as a UML class node, and their compositions, for instance, a UML class node including its field and method descriptions, compose them. 


A \textsf{\textsc{\small{Security Design Solution (SDS)}}} is the aggregation of one or more \ams\ and \ames\ that, taken together, fulfill one or more \srs. This means that any information that is modeled in an \ad\ to fulfill a certain \sr\ either as full \ams\ (e.g., a UML sequence diagram that models a user authentication flow), or as \ames\ in \ams\ (e.g., an ``optional'' flow in a sequence diagram modeling an explicit countermeasure flow within an otherwise non-security-specific sequence diagram) comprises an \sds. All such \sdss\ in an \ad, taken together, comprise the \sad.

To conclude: an \sr\ bridges architectural and security concepts. Since \sdss\ model how \srs\ are fulfilled and \srs\ prescribe \countermeasures, an \sds\ implicitly models one or many \countermeasures, which in turn implies that \ams\ and \ames\ implicitly model security-related information in an \sad. For the remainder, we refer to UML diagrams as \ams\ and to UML modeling elements by \ames.

\section{\uppercase{Methodology}}\label{sec:method}
We followed the objective-centered Design Science Research (DSR) approach described by Peffers et al.~\cite{Peffers2007}, because the underlying problem and its motivation had already been established in~\cite{Sabau2024}. We followed DSR as follows:

First, we define a set of objectives to guide this study, as presented below. Based on these objectives, we develop SA-UML by systematically mapping the security concepts introduced in section~\ref{sec:foundations} onto the UML metamodel. For each concept, we identify suitable UML metaclasses as extension points and analyze their relationships to preserve the original model's semantics. We then demonstrate SA-UML through a case study by documenting structural and behavioral views of a fictitious software system using fixed diagrams as Scalable Vector Graphics (SVGs). We then use the case study component diagrams to conduct an evaluation in the form of a trend analysis with $n=27$ participants. We communicate our work through scientific papers such as this one.

\subsection{Solution Objectives}
The main objective was to avoid maintaining the SAD as an independent artifact and instead integrate security-relevant information into the overall architecture description. To operationalize this objective, we formulated the following objectives:

\begin{enumerate}
    \item \textbf{Conceptual fidelity:} The artifact must represent the central SAM concepts and preserve their intended meaning.
    \item \textbf{UML compatibility:} The artifact should be compatible to standard UML modeling environments such as the Eclipse Modeling Framework (EMF)\footnote{\url{https://eclipse.dev/emf/}} 
    and Papyrus \footnote{\url{https://eclipse.dev/papyrus/}}. 
    
    \item \textbf{Consistency support:} Constraints should validate models and ensure their conformance to the intended semantics.
    \item \textbf{Security as a built-in property:} The profile should make security-relevant information extractable from the models, supporting reasoning about security in architecture models. 
    \item \textbf{Future tool support:} The artifact should provide the basis for a future modeling tool that enables the development of integrated SADs using SA-UML and incorporates the visualization concepts as presented in~\cite{Sabau2024} to support architects in modeling secure architectures.
\end{enumerate}
\section{\uppercase{The SA-UML Profile}}\label{sec:scam-dev}
\begin{figure*}[htb]
\centering
    \includegraphics[width=0.98\linewidth]{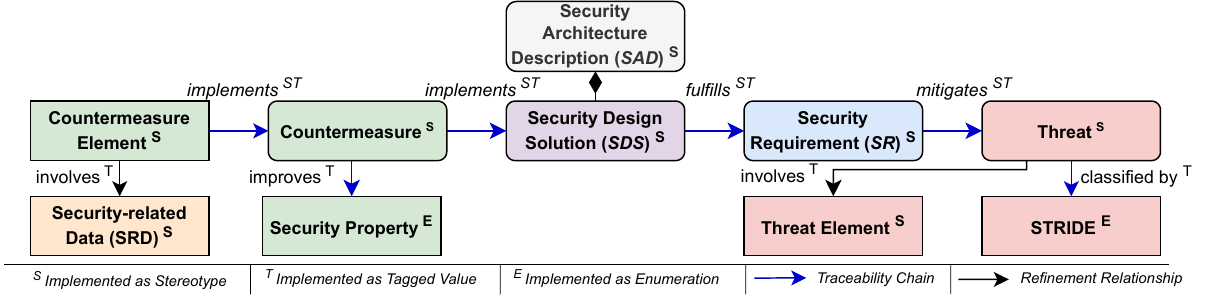}
    \caption{SA-UML solution concept based on SAM (cf. figure~\ref{fig:concept-model}).}
    \label{fig:solution-concept}
\end{figure*}

\subsection{Design of SA-UML}

The conceptual design of SA-UML is depicted in Figure~\ref{fig:solution-concept}. It builds on five SAM concepts: \sad, \sds, \sr, \countermeasure, and \threat. In addition, it introduces six core model-level refinements:

\begin{itemize}
    \item A \countermeasureElement\ is an \ame\ that forms part of a \countermeasure.
    \item \sloppy \textsf{\textsc{\small{Security-related Data (SRD)}}} denotes data whose processing or disclosure is security-relevant.
    \item A \threatElement\ represents an element involved in a \threat.
    \item \stride\ is a common classification of \threats\ into six security categories: \textbf{S}poofing, \textbf{T}ampering, \textbf{R}epudiation, \textbf{I}nformation disclosure, \textbf{D}enial of service, \textbf{E}levation of privilege~\cite{Shostack.2014}.
    \item A \securityProperty\ characterizes a security objective improved by a \countermeasure~\cite{Shostack.2014}.
    \item A \traceabilityChain\ connects concrete \countermeasureElements\ through \countermeasures, \sds, and \srs\ to the \threats\ they ultimately address. The blue arrows in Figure~\ref{fig:solution-concept} represent this chain.
\end{itemize}
The refinements keep security rationale and architectural realization distinct but connected. A \countermeasure\ remains a single security concept even when several \ames\ realize it. \countermeasureElements\ identify these contributions and may involve \srd, thereby making security-relevant data explicit. \threatElements\ anchor a \threat\ in the architecture, while \threatActor\ specializes \threatElement\ to distinguish agency from mere involvement~\cite{NIST.800-63-4}. A \threat\ itself is not an architectural building block because it denotes a harmful condition or scenario rather than something realized by the system. \threats\ are classified using \stride, while \countermeasures\ identify the \securityProperties\ they improve~\cite{Shostack.2014}. These shared vocabularies enable consistent filtering, comparison, and conformance checking. Thus, the four descriptive relationships characterize or decompose individual security concepts rather than connecting them to their architectural realization.

\sds\ and \sr\ bridge the architectural and security levels. Conceptually, an \sds\ represents the architectural model of a \countermeasure\ prescribed by an \sr, while the \countermeasure\ mitigates the \threat. For traceability, SA-UML deliberately projects these semantics onto a single linear chain: it reverses the \sds-to-\countermeasure\ modeling direction and collapses \sr-to-\countermeasure-to-\threat\ into \sr-to-\threat. These choices enable uniform cross-level navigation and tool processing. The direct \sr-to-\threat\ reference is a traceability shortcut; responsibility for mitigation remains with the \countermeasure.

\subsection{Implementation of SA-UML}\label{sub:formal-implementation}
\begin{figure*}[htb]
\centering
    \includegraphics[width=0.98\linewidth]{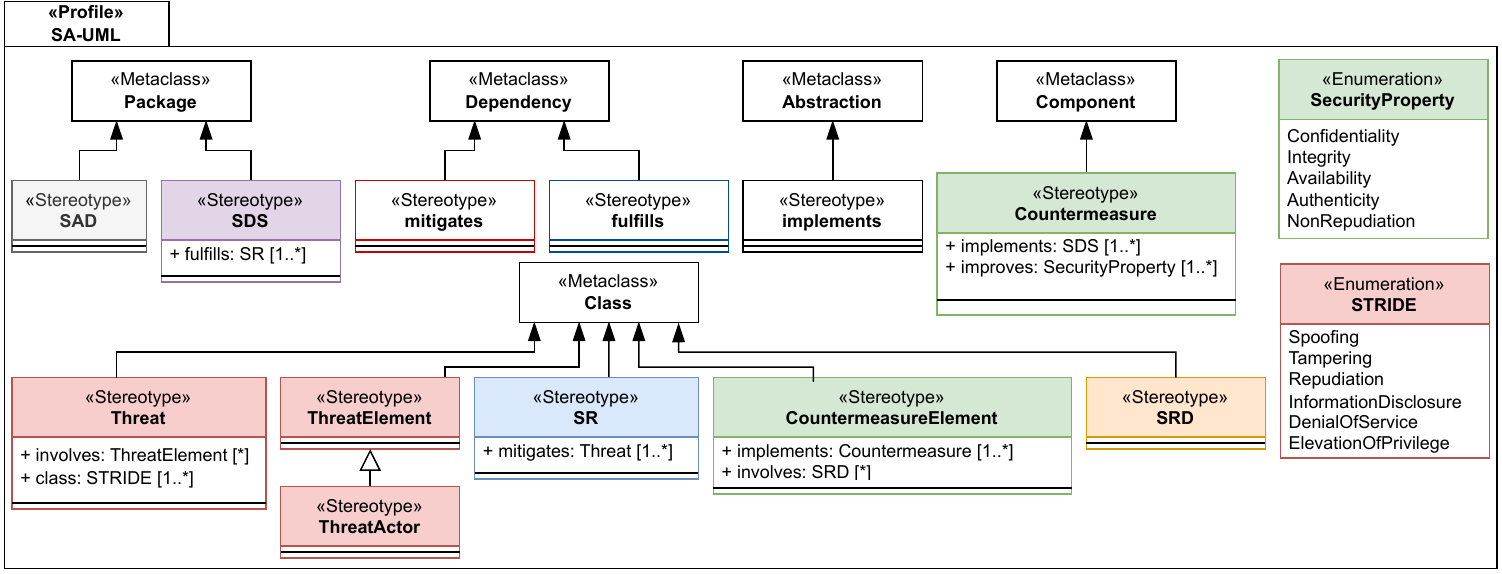}
    \caption{SA-UML profile diagram with the extended \texttt{Stereotypes}, \texttt{Tagged Values}, and \texttt{Enumerations}.}
    \label{fig:sa-uml-profile-diagram}
\end{figure*}

\subsubsection{UML Profile Foundations}

UML profiles are a common approach for extending the UML metamodel~\cite{Jurjens2001} by adding domain-specific semantics without altering the underlying metamodel~\cite{UMLSpecification}.
They provide a UML extension mechanism that can be used with common UML modeling environments, such as Eclipse EMF 
and Papyrus. 

SA-UML is realized through five core UML profile constructs. To distinguish conceptual terminology from its technical implementation, we use \texttt{This Font} for UML concepts; as a reminder, \textsf{\textsc{\small{This Font}}} denotes central terminology at the conceptual level. 
\begin{itemize}
    \item \metaclass: A class in the UML metamodel that defines a type of UML model element, such as \class, \component, or \package. A \stereotype\ references a \metaclass\ as an extension point without modifying it.
    \item \stereotype{}: A domain-specific extension of a UML \metaclass\ that adds semantics to its model elements.
    \item \taggedValue{}: An attribute of a \stereotype{} that stores domain-specific information.
    \item \constraint: A rule used to validate whether a model conforms to the intended profile semantics. \constraints{} are often formulated in the Object Constraint Language (OCL)~\cite{UMLSpecification}.
    \item \enumeration: A fixed set of named literals that can be used as a data type.
\end{itemize}

These constructs allow \ams\ to model both architectural and security-related data.

\subsubsection{Technical Solution}

SA-UML comprises twelve \stereotypes\ extending five UML \metaclasses, eight \taggedValues, two \enumerations, and four OCL constraints.
Figure~\ref{fig:solution-concept} shows how the SAM concepts and SA-UML refinements are represented at a high level. S, T, and E denote implementation as a \stereotype, \taggedValue, and \enumeration, respectively. ST denotes a \taggedValue\ with an optional relational \stereotype.


Figure~\ref{fig:sa-uml-profile-diagram} shows the lower-level implementation of SA-UML as a UML profile diagram. For brevity, the OCL constraints are not depicted but are presented in Section~\ref{sub:tagged-values-and-tracing}. The complete SA-UML profile is available in the accompanying Zenodo artifact.

For each concept, we selected the UML \metaclass\ that best fit semantically. This selection considered the UML meaning of the \metaclass, the concept's purpose, and the relationships it must maintain with other concepts.

The scoping concepts, \sad\ and \sds, extend \package. Both concepts define architectural boundaries rather than typeable model content: the \sad\ frames the security-relevant view of the entire architecture, while an \sds\ groups the \ames\ that collectively fulfill one or more \srs. The grouping and namespace semantics of \package\ therefore correspond to both concepts' roles.
An \sad\ aggregates its \sds\ packages through standard UML package nesting: the outer \sad\ package contains the \sds\ packages belonging to the system's security architecture.

Unlike these scoping concepts, a \countermeasure\ represents a security mechanism realized by concrete model elements. It therefore extends \component, whose semantics capture a modular architectural element with concrete realizations.

The remaining concept-related \stereotypes, namely \threat, \threatElement, \sr, \countermeasureElement, and \srd, extend \class\ because they require independently identifiable, typeable, and referenceable model elements. This is particularly relevant for an \sr, which must connect a \threat\ to its architectural solution, and for \srd, which must participate in the UML type system. The \threatActor\ \stereotype\ specializes \threatElement\ and therefore inherits its extension of \class. The generalization capabilities of \class\ also support the hierarchical organization of these concepts.

After mapping the concept-related elements, the remaining extensions formalize their relationships. The relationships \mitigates\ and \fulfills\ extend \dependency, whereas \implements\ extends \abstraction\ because it connects elements at different levels of abstraction.

\subsection{Traceability Chain and Tagged Values}\label{sub:tagged-values-and-tracing}
The \traceabilityChain\ enables bidirectional navigation between security elements and their architectural implementations through \taggedValues.
Because the traceability information is stored in the semantic UML model, its interpretation does not depend on particular diagrams. A modeling tool can use this information to derive focused views by hiding unrelated elements while leaving the underlying \ad\ unchanged.

The \traceabilityChain\ is encoded through six typed references:
\begin{itemize}
    \item \texttt{CountermeasureElement.implements: Countermeasure [1..*]},
    \item \texttt{Countermeasure.implements: SDS [1..*]},
    \item \texttt{Countermeasure.improves: SecurityProperty [1..*]},
    \item \texttt{SDS.fulfills: SR [1..*]}, 
    \item \texttt{SR.mitigates: Threat [1..*]}, and
    \item \texttt{Threat.class: STRIDE [1..*]}.
\end{itemize}
A profile-conformant UML validator automatically checks the lower bound \texttt{[1..*]} on each trace reference. This lower bound requires every source element to reference at least one target, thereby preventing incomplete chains without an additional OCL invariant.
The optional \texttt{involves} properties further characterize individual security concepts.
References are stored only in the direction shown in Figure~\ref{fig:solution-concept}. Reverse navigation is derived by querying the referencing elements rather than maintaining redundant inverse references.

Relational \stereotypes\ provide optional graphical representations of the machine-readable \taggedValues.
OCL is required only to prevent the two representations from diverging by ensuring that every supplier of a stereotyped dependency occurs in the corresponding \taggedValue\ of its client.
The following invariant illustrates this rule for \mitigates:
\begin{lstlisting}[
  style=ocl,
  caption={Representative OCL consistency constraint.},
  label={lst:ocl-consistency}
]
context SR
inv singleSourceOfTruth:
  self.base_Class.clientDependency
  ->select(d | d.getAppliedStereotypes()
    ->exists(s | s.name = 'mitigates'))
  ->forAll(d | d.supplier->forAll(t |
    self.mitigates.base_Class->includes(t)))
\end{lstlisting}

\subsection{Diagram Support}\label{sub:diagram-support}
SA-UML can be used in structural and behavioral diagrams.

\subsubsection{Structural Diagrams}

\paragraph{Package diagrams:} Provide an architectural overview, with \sad\ framing the overall \sa\ and \sds\ partitioning it into different domains. This enables several security-specific views of the overall \ad.

\paragraph{Class diagrams:} Capture the static security ontology through classes carrying the \threat, \threatElement, \sr, \countermeasureElement, and \srd\ stereotypes together with their relationships.

\paragraph{Component diagrams:} Document an abstract view of the software by modeling components and their interfaces rather than their internal composition.
Because \component\ specializes \class, \stereotypes\ extending \class\ can also be applied to \component. \countermeasure\ is deliberately restricted to \component\ because it denotes a security mechanism that must be implemented as an architectural component. Applying it to a plain \class\ is therefore outside the profile's intended scope.
An \sds\ groups the \countermeasures\ that collectively implement the solution and thereby support the fulfillment of one or more \srs. The relationships between security concepts and architectural elements follow the same structure as in class diagrams. Classes carrying \srd\ participate naturally in the UML type system by serving as attribute and operation-parameter types.

\subsubsection{Behavioral Diagrams}

Security semantics in behavioral diagrams are carried exclusively through the UML type system. Security-relevant parameters and attributes are typed by classifiers carrying \srd\ rather than by applying \stereotypes\ directly to messages or properties. This approach maintains a single annotation point at the classifier level and avoids redundant or inconsistent labeling across diagram types.

\paragraph{Sequence diagrams:} Model security-relevant communication flows through lifelines whose represented elements are typed by classifiers carrying \stereotypes\ that extend \class. Classes carrying \srd\ can serve as operation-parameter types to express data sensitivity within a message.

\paragraph{Activity diagrams:} Support process-level modeling through object nodes typed by classifiers carrying an applicable security stereotype, thereby enabling sensitive data flows to be traced.

\section{\uppercase{Demonstration}}\label{sec:demonstration}
We designed a fictitious web-based document portal and modeled it in SA-UML to demonstrate SA-UML in a case study. The portal is designed to allow users to log in, upload, and access confidential documents. Because the \ad\ of the case study also served as the \ad\ used in the evaluation, we kept its design simple; i.e., we aimed to demonstrate the core SA-UML features without unnecessary complexity. 

We formulated the portal around a set of functional and security requirements~\ref{tab:systemRequirements}. To ensure UML conformance, we first designed the \ams\ in EMF. After ensuring conformance, we recreated them in draw.io\footnote{\url{https://app.diagrams.net/}}. Here, we applied custom stereotype styling. Profile semantics remain encoded by stereotype keywords, UML shapes, and labeled relationships. 

\begin{figure*}[htb]
\centering
    \includegraphics[width=\textwidth]{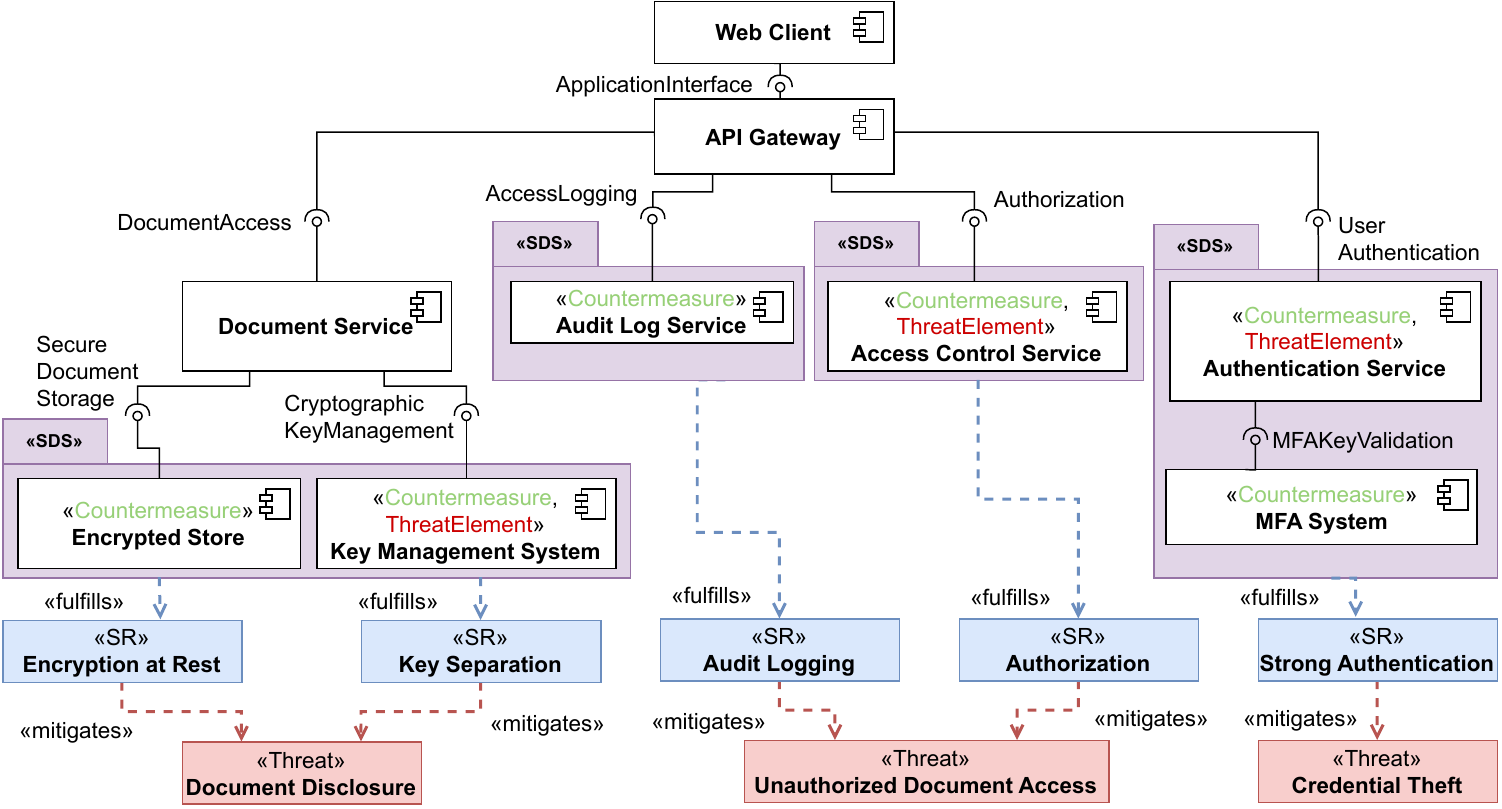}
    \caption{UML component diagram of case study in SA-UML.}
    \label{fig:sa-uml-component-diagram}
\end{figure*}

\begin{figure*}[htb]
\centering
    \includegraphics[width=\textwidth]{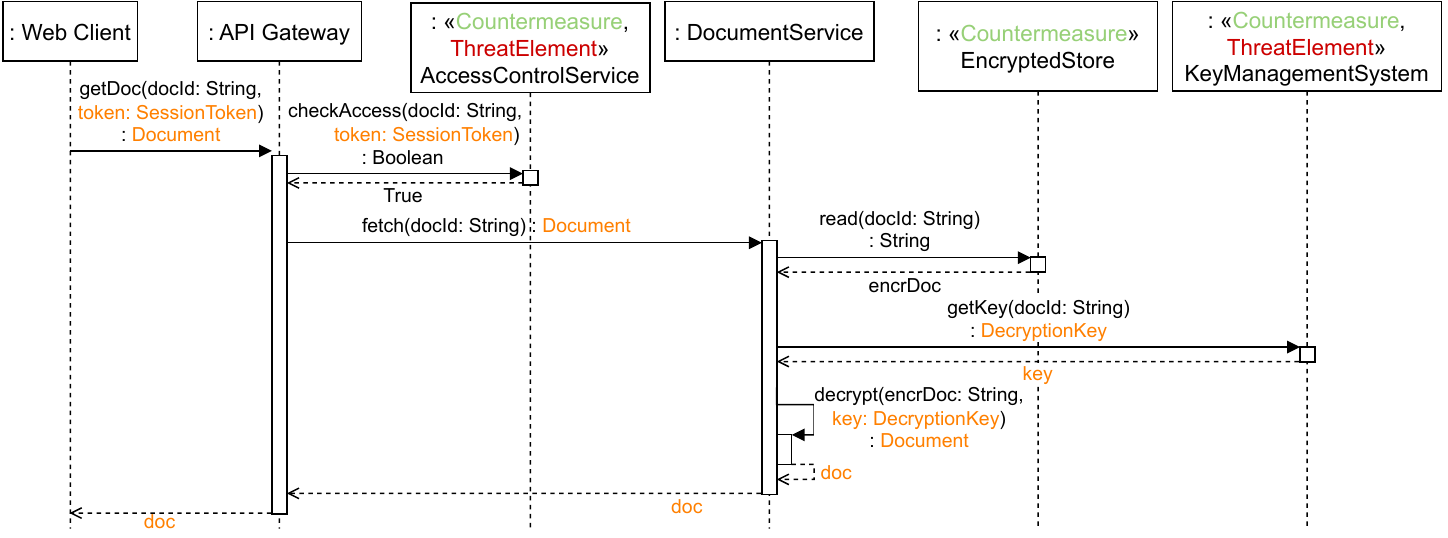}
    \caption{SA-UML sequence diagram for confidential document retrieval.}
    \label{fig:sa-uml-sequence-diagram}
\end{figure*}

We designed a component diagram to model the web portal's structure as a structural view (Figure~\ref{fig:sa-uml-component-diagram}) and a sequence diagram of a use case to model its behavior (Figure~\ref{fig:sa-uml-sequence-diagram}). We kept both diagrams simple for analysis. For example, the diagrams omit \countermeasureElements; the sequence diagram omits the authentication flow and the \textit{Audit Log Service}.


\begin{table}[!htb]
\footnotesize
\caption{Functional and security requirements of the case study software system.}
\label{tab:systemRequirements}

\renewcommand{\arraystretch}{1.15}
\begin{tabularx}{\linewidth}{L{0.2\linewidth} Y}
\toprule
\rowcolor{white}
\textbf{Type of Requirement} & \textbf{Requirements} \\
\midrule
\textbf{Functional} & [FR-1] Users can log in to the system.\newline
     [FR-2] Documents can be uploaded and downloaded.\newline
     [FR-3] Access to documents is restricted to authorized users. \\
\midrule
\textbf{Security} & [SR-1] Users must be able to authenticate to the system.\newline
    [SR-2] The system must ensure access is granted only to authorized users.\newline
    [SR-3] The system must use an encrypted storage for any persistent documents.\newline
    [SR-4] Cryptographic keys and stored data must be separated.\newline
    [SR-5] The system must audit any document accesses.\\
\bottomrule
\end{tabularx}
\end{table}


\subsection{Structural View}
The document portal is designed as a microservice-based system comprising nine services. Each service addresses a functional or security-related concern, as summarized in Table~\ref{tab:systemRequirements}, and is modeled as a UML component in Figure~\ref{fig:sa-uml-component-diagram}.

Users log in and access documents through the \textit{Web Client}. The \textit{API Gateway} routes all Web Client requests to the individual services. The \textit{Document Service} handles document upload, retrieval, deletion, encryption, and decryption. [SR-3] and [SR-4] prescribe secure document storage. The \textit{Encrypted Store} and the \textit{Key Management System} implement this by storing and managing the keys used to encrypt the documents. [SR-1], [SR-2], and [SR-5] are realized by the \textit{Authentication Service}, Multi-Factor Authentication System (\textit{MFA System}), and \textit{Audit Log Service}.

The \ames\ that belong to the \sad\ are identified by color: The red \ames\ model \threats\ that the \countermeasures\ aim to mitigate; the blue \ames\ model the corresponding \srs\ between the services and the \threats; the purple \ames\ model \sdss\ as packages, grouping the components that implement each \sds\ as \countermeasures\ to the corresponding \threats. Lastly, in accordance with \texttt{SDS.fulfills: SR}, the \fulfills\ dependencies originate from these packages rather than from the contained components. Some services are annotated as a \threatElement\, indicating potential attack surfaces within the architecture.


\subsection{Behavioral View}
Figure~\ref{fig:sa-uml-sequence-diagram} shows a sequence diagram in SA-UML for the use case of a user retrieving a confidential document. The model follows the component diagram presented earlier. Six main actors realize this message sequence, modeled as standard UML sequence-diagram object nodes. As above, the object roles are labeled to indicate the security-related role as a \countermeasure\ or a \threatElement. Moreover, orange text indicates \srd: \texttt{token: SessionToken} is used for access control; \texttt{key: DecryptionKey} is used for data document encryption; \texttt{doc: Document} represents the decrypted confidential document that was accessed through the web client.

Overall, this case study shows how an \sad\ can be an integrated but extractable part of an \ad\ and how this integrated modeling approach applies to SA-UML. 

\section{\uppercase{Evaluation}}~\label{sec:evaluationSetup}

We conducted a questionnaire study to gather initial indications as to whether SA-UML generally helps in understanding and tracing security-relevant architectural information. The evaluation was grounded in three evaluation questions:

\begin{enumerate}
    \item How does SA-UML affect efficiency?
    \item How does SA-UML affect effectiveness?
    \item How is SA-UML rated?
\end{enumerate}

\subsection{Evaluation Setup}

\subsubsection{Participant Selection}
The questionnaire targeted computer science students and professionals with backgrounds in software engineering, software architecture, UML, and, optionally, security. We chose this target group as they reflect the main audience of SA-UML: non-security architects and engineers, who are involved in security-related architecture tasks. 
We used purposive sampling with snowball sampling, using mailing lists, LinkedIn, and personal contacts~\cite{Baltes2022}.

\subsubsection{Questionnaire Design}\label{subsub:eval-questionnaire-design}
The study used a within-subject design: each participant completed tasks for both a standard UML component diagram and an SA-UML component diagram of the system introduced in Section~\ref{sec:demonstration}. Presentation order was randomized, yielding two sequence groups: Group A encountered UML first and SA-UML second; Group B encountered SA-UML first and UML second. Both groups therefore evaluated both modeling approaches.

We presented the diagrams through an SVG, HTML, and JavaScript interface, allowing participants to interact with the models. For SA-UML models, participants could switch between standard UML and SA-UML views to compare the models.

The survey followed a task-based design: one single-choice and four multiple-choice questions were asked per diagram. The questions were designed to operationalize the three evaluation questions. Efficiency was measured through page dwell time to answer the questions regarding the diagram. Effectiveness was assessed through the correctness of responses, providing specific insights into 1) the understanding of the underlying architecture, 2) the ability to trace relationships between architectural elements and security requirements, and 3) the security understanding.

Lastly, we asked seven Likert-scale questions to assess different aspects of participants' perceptions of SA-UML in solving the tasks, including an optional free-text rationale field.
Table~\ref{tab:likert-items} provides an overview of the Likert questions; the full questionnaires are published as artifacts~\cite{zenodo}.

\subsubsection{Evaluation Hypotheses}
We formulated five hypotheses to address the three evaluation questions and explore whether performance changed between rounds. The first three concern efficiency, effectiveness, and perceived usefulness. We expected explicit security information and traceability links to reduce search effort, support correct answers, and be perceived as useful:

\begin{itemize}
    \item[] \textbf{H1 (efficiency)}: Participants complete the five tasks in less total time with SA-UML than with standard UML.
    \item[] \textbf{H2 (effectiveness)}: Participants achieve higher response scores with SA-UML than with standard UML.
    \item[] \textbf{H3 (perceived usefulness)}: Ratings of SA-UML exceed the neutral midpoint of 3 for each evaluated item or scale, after reverse-coding the negatively worded item.
\end{itemize}

Completion time is measured in seconds using page-level timing; response scores range from $0$ to $1$. H1 and H2 compare the two conditions within participants with a positive value indicating better results with SA-UML. H3 evaluates the individual overall and architecture-related items and the combined security-related scale listed in Table~\ref{tab:likert-items}. These comparisons concern the presented conditions, which differed in security-information content (Section~\ref{subsub:questionnaire-design}).

Because participants encountered the same system twice, familiarity could also improve second-round performance. We therefore formulated two additional, exploratory hypotheses to support the interpretation of H1 and H2:

\begin{itemize}
    \item[] \textbf{H4 (round-related efficiency)}: Participants complete the tasks faster in round 2 than in round 1.
    \item[] \textbf{H5 (round-related effectiveness)}: Participants achieve higher response scores in round 2 than in round 1.
\end{itemize}

All five hypotheses were tested one-tailed in the stated directions. H4 and H5 were tested separately in Groups A and B. These comparisons examine second-round improvement within each presentation sequence; they do not isolate learning effects from differences between the modeling conditions.
\begin{table}[!tb]
\small
\caption{Overview of Likert-scale items. O-item: Overall; A-items: Architecture-related; S-items: Security-related}
\label{tab:likert-items}

\renewcommand{\arraystretch}{1.15}
\rowcolors{2}{rowgray}{white}
\begin{tabularx}{\linewidth}{L{0.05\linewidth} Y}
\toprule
\rowcolor{white}
\textbf{ID} &\textbf{Item} \\
\midrule
$\text{O}$ & SA-UML provides a general advantage in larger project/industry.  \\
$\text{A}1$ & The additional elements perceived as distracting. \\
$\text{A}2$ & SA-UML helped participants understand the system architecture. \\
$\text{A}3$ & The integration of security aspects was comprehensible. \\
$\text{S}1$ & SA-UML helped participants understand threats at the architectural level.  \\
$\text{S}2$ & SA-UML helped participants understand countermeasures and security solutions. \\
$\text{S}3$ & SA-UML helped participants understand the relationships between security requirements and solutions. \\
\bottomrule
\end{tabularx}
\end{table}

\subsubsection{Data Preparation}
A total of $n = 63$ people participated in the survey, of whom $n = 30$ completed it.
We decided to remove two data sets with unit non-responses because they do not contribute any real value to the data collection.
Additionally, a data set in which SA-UML was not used was excluded from the analysis because it cannot support a comparison between standard UML and SA-UML modeling.
This resulted in a data set of $n = 27$, which was used for the data analysis~\cite{Wohlin.2000}.

\subsection{Evaluation Results}\label{sec:evaluation}


\subsubsection{Demographics}\label{sub:evaluation:demographic}
\begin{table}[!htb]
\centering
\small
\caption{Participant prior experience ($n=27$)}
\label{tab:demographic-information}

\renewcommand{\arraystretch}{1.15}
\setlength{\tabcolsep}{3pt}
\rowcolors{2}{rowgray}{white}

\begin{tabularx}{\linewidth}{L{0.36\linewidth} Y Y Y Y}
\toprule
\rowcolor{white}
& \multicolumn{2}{c}{\textbf{Professionals (n=19)}}
& \multicolumn{2}{c}{\textbf{Students (n=8)}} \\
\rowcolor{white}
\cmidrule(lr){2-3}
\cmidrule(lr){4-5}
\rowcolor{white}
\textbf{Characteristic}
& \textbf{Count} & \textbf{\%}
& \textbf{Count} & \textbf{\%} \\

\midrule
\rowcolor{white}
\multicolumn{5}{l}{\textbf{Prior UML experience}} \\
None
& 2 & 10.5 & 0 & 0.0 \\
Basic / theoretical
& 4 & 21.1 & 2 & 25.0 \\
Occasional / practical
& 8 & 42.1 & 2 & 25.0 \\
Regular / projects
& 5 & 26.3 & 2 & 25.0 \\
Work focus / out-of-school
& 0 & 0.0 & 2 & 25.0 \\

\midrule
\rowcolor{white}
\multicolumn{5}{l}{\textbf{Prior security experience}} \\
None
& 1 & 5.3 & 1 & 12.5 \\
Basic / theoretical
& 0 & 0.0 & 1 & 12.5 \\
Occasional / practical
& 6 & 31.6 & 1 & 12.5 \\
Regular / projects
& 11 & 57.9 & 3 & 37.5 \\
Work focus / out-of-school
& 1 & 5.3 & 2 & 25.0 \\

\bottomrule
\end{tabularx}

\smallskip
\parbox{\linewidth}{%
\textit{Note.} Percentages refer to the respective subgroup
and may not sum to 100 due to rounding.
}
\end{table}
The dataset consists of 27 fully completed questionnaires, comprising 19 working professionals and eight students.

Overall, professionals have a high level of formal education and domain experience: most hold a master's degree or diploma ($63.2\,\%)$. 
Three professionals have a PhD and three professionals hold a bachelor's degree (each $15.8\,\%$).
The majority work in IT or software development ($63.2\,\%$) and report more than ten years of professional experience ($63.2\,\%$).
Four of the seven professionals who do not work in IT work in education or research ($21.1\,\%$).
The student subsample is predominantly enrolled in computer science or business informatics master's programs ($87.5\,\%$).

Prior experience with UML and security is shown in Table~\ref{tab:demographic-information}. Professionals report substantially more
practical experience, especially in security-related work, whereas students' self-assessments are more evenly distributed across the available experience levels.



\subsubsection{Hypotheses Test Results}\label{sub:evaluation:result}
SoSciSurvey tracked the time spent on each page.
In addition to the time data, we analyzed responses to the questions on the component diagrams by calculating a score between $0$ and $1$ for each question.

For single-choice questions:
\begin{equation}
Score =
\begin{cases}
1, & \text{if the answer is correct} \\
0, & \text{otherwise}
\end{cases}
\end{equation}

For multiple-choice questions:
\begin{equation}
    Score = \frac{TP + TN}{n_{\text{scored}}}
\end{equation}
where $TP$ denotes correctly selected options, 
$TN$ correctly rejected options, 
and $n_{\text{scored}}$ all scored answer options, 
excluding optional answers.

\paragraph{Completion Time.} Participants completed the five tasks in a mean of $247.50$ seconds using UML and $169.80$ seconds using SA-UML. The paired difference $T_{\mathrm{UML}} - T_{\mathrm{SA\text{-}UML}}$ had a mean of $77.67$ seconds ($Mdn = 76.00$, $SD = 143.49$), indicating shorter recorded completion times with SA-UML in this sample.

The Shapiro--Wilk test did not provide evidence against the normality of the paired differences ($W = 0.9621$, $p = 0.4130$), and the corresponding Q--Q plot showed no substantial deviation from normality. An exploratory one-tailed paired $t$-test yielded $t(26) = 2.81$, $p = 0.005$, and $d_z = 0.54$ ~\cite{field.2018,Lakens.2013}. Thus, the results provide an initial indication that participants required less time to analyze the SA-UML diagram than the corresponding UML diagram. Because the recorded measure represents page-level completion time and the study has an exploratory design, the result should not be generalized beyond the present sample.

\paragraph{Response Scores.}
Mean response scores were $0.814$ ($SD = 0.096$) for UML and $0.835$ ($SD = 0.096$) for SA-UML. The paired difference $S_{\mathrm{SA\text{-}UML}} - S_{\mathrm{UML}}$ had a mean of $0.020$ ($SD = 0.070$).

The Shapiro--Wilk test did not provide evidence against the normality of the paired differences ($W = 0.9517$, $p = 0.2359$), and the corresponding graphical diagnostics showed no substantial deviation from normality. An exploratory one-tailed paired $t$-test yielded $t(26) = 1.50$, $p = 0.072$, and $d_z = 0.29$ ~\cite{field.2018,Lakens.2013}. The observed score difference was therefore inconclusive: the analysis provides neither clear evidence of higher response accuracy with SA-UML nor evidence that both approaches are equivalent. Because the descriptive differences varied by presentation sequence, possible period and order effects are considered separately in Section~\ref{sec:discussion}.

\paragraph{Perceived Usefulness.}
The options on the Likert scale ranged from strongly disagree ($1$) to strongly agree ($5$).
Since item $\text{A}1$ is negatively worded (cf. Table~\ref{tab:likert-items}), the results were reverse-coded for the analysis.
For each A and S item set in Table~\ref{tab:likert-items}, we calculated Cronbach's $\alpha$ to measure scale reliability.
Since Cronbach's $\alpha$ reached an acceptable level only for the S-items ($\alpha = 0.814 \geq 0.8$)~\cite{field.2018}, they were aggregated into a scale; A-items were evaluated individually.

All Shapiro-Wilk tests on the answer distributions resulted in a p-value of $p<0.01$~\cite{shapiro.1965}.
Thus, we conducted Wilcoxon signed-rank tests against the neutral value of three ($3$)~\cite{field.2018}.
The results are shown in Table~\ref{tab:wilcoxon-likert}.
\begin{table}[!tb]
\small
\caption{Results of Wilcoxon Signed Rank test on Likert--items.}
\label{tab:wilcoxon-likert}

\renewcommand{\arraystretch}{1.15}
\rowcolors{2}{rowgray}{white}
\begin{tabularx}{\linewidth}{L{0.21\linewidth} Y Y Y Y Y Y}
\toprule
\rowcolor{white}
\textbf{Item} &\textbf{M} & \textbf{Mdn} & \textbf{W} & \textbf{r} & \textbf{z} & \textbf{p $<$} \\
\midrule
$\text{O}$ & $4.13$ & $4.00$ & $235.00$ & $0.78$ & $3.61$ & $0.001$ \\
$\text{A}1$ & $4.41$ & $5.00$ & $365.00$ & $0.84$ & $4.36$ & $0.001$ \\
$\text{A}2$ & $3.82$ & $4.00$ & $208.50$ & $0.73$ & $3.37$ & $0.001$ \\
$\text{A}3$ & $4.37$ & $5.00$ & $276.00$ & $0.91$ & $4.34$ & $0.001$ \\
$\text{S}1$ to $\text{S}3$ (scale) & $4.09$ & $4.33$ & $283.00$ & $0.78$ & $3.82$ & $0.001$ \\
\bottomrule
\end{tabularx}
\end{table}
The effect size of $r>0.5$ and the p-value of $p<0.001$ for all items, implies that SA-UML was perceived as useful with a large effect size ~\cite{Wohlin.2000,field.2018}.

\paragraph{Round and Presentation-Order Trends:}
To explore whether the results indicate a general improvement in the second round, the round differences were examined separately for both presentation sequences.
Table~\ref{tab:task-order} shows the distribution of participants across these sequences.
\begin{table}[!h]
\small
\caption{Participant distribution of the task-order.}
\label{tab:task-order}

\renewcommand{\arraystretch}{1.15}
\rowcolors{2}{rowgray}{white}
\begin{tabularx}{\linewidth}{L{0.3\linewidth} Y Y Y}
\toprule
\rowcolor{white}
\textbf{Sequence Group} & \textbf{Overall} & \textbf{Professionals} & \textbf{Students} \\
\midrule
A: UML $\rightarrow$ SA-UML & $17$ & $12$ & $5$ \\
B: SA-UML $\rightarrow$ UML & $10$ & $7$ & $3$ \\
\bottomrule
\end{tabularx}
\end{table}

Table~\ref{tab:diff-and-t-test-round} reports the descriptive differences and the corresponding one-tailed paired t-tests.
\begin{table*}[!tb]
\small
\caption{Differences and paired t-test results on completion time and response scores. \texttt{$T_{Rx}$} = time captured for completing all tasks depending on round $x$. \texttt{$S_{Rx}$} = score achieved depending on round $x$. Interpretation: positive time difference = second round faster; positive score = second round achieved higher score.}
\label{tab:diff-and-t-test-round}
\renewcommand{\arraystretch}{1.3}
\rowcolors{2}{rowgray}{white}
\begin{tabularx}{\linewidth}{L{0.2\linewidth} Y Y Y Y Y Y}
\toprule
\rowcolor{white}

  & \multicolumn{3}{c}{\textbf{Group A: UML $\rightarrow$ SA-UML}}
  & \multicolumn{3}{c}{\textbf{Group B: SA-UML $\rightarrow$ UML}} \\
\rowcolor{white}
\midrule
\textbf{Difference}
  & \textit{Mean} & \textit{Median} & \textit{SD}
  & \textit{Mean} & \textit{Median} & \textit{SD} \\
  \midrule
$\mathit{T}_{\text{R1}} - \mathit{T}_{\text{R2}}\ (s)$
  & $37.53$ & $51.00$    & $131.66$
  & $-145.90$ & $-151.50$ & $142.99$ \\
$\mathit{S}_{\text{R2}} - \mathit{S}_{\text{R1}}$
  & $0.04$  & $0.04$  & $0.08$
  & $0.02$  & $0.00$  & $0.04$ \\
  \midrule
  \midrule
  \textbf{Paired t-test result}
  & \textit{$d_z$} & \textit{p} & \textit{Significance}
  & \textit{$d_z$} & \textit{p} & \textit{Significance} \\
  \midrule
  Time & $0.28$ & $0.129$ & not significant & $-1.02$ & $0.995$ & not significant\\
  Score & $0.54$ & $0.02$ & significant & $0.37$ & $0.138$ & not significant\\
\bottomrule
\end{tabularx}
\end{table*}

Before applying the tests, we inspected the difference scores graphically and assessed them using the Shapiro-Wilk test~\cite{shapiro.1965,field.2018}. In Group A, response scores were higher in the second round with SA-UML ($M_{\Delta}=0.04$, $d_z=0.54$, $p=0.020$). Additionally, task completion time in the second round with SA-UML was shorter ($M_{\Delta}=37.53$\,s, $d_z=0.28$, $p=0.129$). However, the time-difference results did not meet the $\alpha = 0.05$ threshold.
In Group B, the score difference was smaller ($M_{\Delta}=0.02$, $d_z=0.37$, $p=0.138$), and completion time was longer in the second round with UML ($M_{\Delta}=-145.90$\,s, $d_z=-1.02$, $p=0.995$ for the hypothesized direction).
Thus, only one of the four exploratory comparisons crossed the significance threshold of $\alpha = 0.05$, providing limited evidence for a general second-round improvement. Since modeling approach and round are coupled within each presentation sequence, these results describe order-dependent trends rather than isolating a learning effect.

\begin{table}[!b]
\small
\caption{Descriptive statistics by modeling approach.}
\label{tab:task-time-scores}

\renewcommand{\arraystretch}{1.15}
\rowcolors{2}{rowgray}{white}
\begin{tabularx}{\linewidth}{L{0.2\linewidth} Y Y Y Y Y}
\toprule
\rowcolor{white}
\ &\textbf{Mean} & \textbf{Median} & \textbf{SD} & \textbf{Min} & \textbf{Max}\\
\midrule
 &  \multicolumn{5}{c}{\textbf{Completion Time (s)}} \\
UML & $247.50$ & $219.00$ & $120.01$ & $114.00$ & $561.00$ \\
SA-UML & $169.8$ & $144.00$ & $82.49$ & $52.00$ & $355.00$ \\
\midrule
\midrule
 &  \multicolumn{5}{c}{\textbf{Scores}} \\
 UML & $0.814$ & $0.854$ & $0.096$ & $0.622$ & $0.956$ \\
 SA-UML & $0.835$ & $0.854$ & $0.096$ & $0.648$ & $0.956$ \\
\bottomrule
\end{tabularx}
\end{table}

Table~\ref{tab:task-time-scores} summarizes the completion times and response scores observed for both modeling approaches.
\section{\uppercase{Discussion and Limitations}}\label{sec:discussion}
Below, we interpret the evaluation results, discuss threats to validity, and outline the limitations. 

\subsection{Result Interpretation}

\subsubsection{Efficiency (Time)}
The time measurements indicate a preliminary efficiency trend. Participants completed the tasks on average $77.67$ seconds faster in the SA-UML condition, requiring approximately $30\,\%$ less time than with standard UML ($\frac{time_\text{SA-UML}}{time_\text{UML}} = 0.686$). Comments suggest that color coding (Cases 343, 359) and direct traceability (Cases 343, 356) reduced mental mapping and search effort.

However, the conditions were not information-equivalent: the SA-UML diagram contained explicit SRs and threats absent from standard UML, while the security-focused tasks queried this information. The difference therefore combines information availability, visual encoding, and traceability and cannot be attributed to the notation alone. Although roughly $78$ seconds could be relevant in time-constrained reviews, this requires confirmation against an information-equivalent baseline. The limitation on the questionnaire design is further discussed in Section~\ref{subsub:questionnaire-design}.
Moreover, SoSciSurvey measured page dwell time rather than active interaction, and pauses or distractions may have contributed to the high variance ($SD = 143.49$ seconds). As no general second-round advantage was confirmed, fatigue, breaks, and reviewing previous answers also remain possible influences.

\subsubsection{Effectiveness (Score)}
SA-UML did not significantly increase response scores; its effect on task performance therefore remains inconclusive. Three factors may have limited the observable difference. First, the deliberately simple example produced a ceiling effect. Participants noted that basic experience was sufficient to infer security responsibilities (Case 357) and expected greater benefits for larger systems (Cases 336, 343, 353). Second, scores improved significantly only when SA-UML appeared second, indicating an order dependency.

Third, architectural diagrams permit legitimate alternative interpretations that the scoring metric cannot fully capture. Optional answers accommodated some variation, but one participant still questioned whether the \textit{TLS} connection to the \textit{API Gateway} was security-relevant (Case 346), supporting our treatment of the \textit{API Gateway} and \textit{Web Client} as optional in some questions. Thus, ceiling effects, presentation order, and score subjectivity may explain the small difference, but the data neither demonstrate nor rule out improved effectiveness.

\subsubsection{Perceived Usefulness (Likert)}
Perceived usefulness showed the clearest positive pattern. All Likert items were significantly above the neutral midpoint with large effect sizes, supported by the qualitative responses. Participants particularly valued traceability between SRs, threats, and other components (Cases 343, 344, 356, 394) and expected benefits for less experienced developers or those without a security background (Cases 353, 388).

Responses suggest that SA-UML primarily supports security rather than general architectural understanding (Cases 343, 344), consistent with its intended scope. The optional security overlay was also viewed positively (Case 346). Conversely, one participant considered the additional elements redundant in the small example and questioned the completeness of modeled threats (Case 378). Overall, the responses indicate perceived value and easier access to security information, but not an objectively demonstrated improvement in understanding.

\subsection{Learning and Ordering Effects}
Despite randomization, the secondary analyses revealed order-dependent patterns. Only one of four comparisons was significant: Group A scored higher in the subsequent SA-UML round, possibly because familiarity reduced cognitive overhead while SA-UML made security information easier to locate. Conversely, Group B required on average more than two additional minutes for the subsequent UML tasks, possibly because they expected similar security cues. As the reverse effect was not significant and the patterns were inconsistent, the analyses establish no general ordering effect. However, learning and contrast effects remain threats to internal validity.




\subsection{Threats to Validity}

\subsubsection{Conclusion Validity}
The study's statistical power is limited by the sample size, particularly for detecting small effects. A larger sample might have yielded a significant result for this hypothesis.


\subsubsection{Internal Validity}
As discussed above, the most significant threat to internal validity is the potential for learning and ordering effects. A second threat concerns the subjectivity of the scoring metric, as architectural diagrams allow multiple legitimate interpretations that the scoring formula may not fully capture. We minimized both threats by randomizing task order and providing optional answers.


\subsubsection{Construct Validity}

The simple example diagram introduces a ceiling effect. Moreover, SoSciSurvey's page-level time tracking cannot distinguish active task engagement from unrelated activity. With both UML and SA-UML yielding high scores, the score metric lacked sufficient sensitivity to detect expected differences.


\subsubsection{External Validity}

The study's external validity is limited in several ways, which we acknowledge, given the goal of a trend analysis evaluation to gather early, fast feedback. With $n = 27$ non-randomly sampled participants, the results cannot be considered statistically representative of, or generalized to, the broader population of software architects or security engineers. Moreover, the sample is dominated by participants with an IT or research background and advanced degrees, which may not reflect the diversity of practitioners who would use SA-UML in practice~\cite{Baltes2022}. Furthermore, the study used only a single diagram type and a single, intentionally simple example system, leaving open whether the observed benefits extend to other diagram types or larger systems—a limitation participants themselves noted (Cases 334, 349, 378). Finally, the controlled setting lacks the complexity of real-world development contexts, including team collaboration, domain knowledge, and time pressure.



\subsection{Limitations}\label{sub:limitations}
This study is subject to limitations concerning its questionnaire design, prototype implementation, and evaluation scope.

\subsubsection{Questionnaire Design}\label{subsub:questionnaire-design}
The questionnaire design introduced a confound because the two conditions of the standard UML diagram and SA-UML diagram were not information-equivalent. The SA-UML diagram explicitly contained \srs\ and \threats\ that were absent from the standard UML diagram, while the security-focused tasks directly queried this content. Consequently, the observed $77.67$-second difference conflates information availability with visual encoding and cannot be attributed to the notation alone.
Because the evaluated rendering combined stereotype labels, UML shapes, and color highlighting, the study neither isolates the effect of color nor evaluates accessibility under greyscale or color-vision deficiency. A fair baseline should provide identical security content through standard UML notes or an accompanying \sad\ containing the same \srs, \threats, and mappings. 

\subsubsection{Evaluation Scope}
The evaluation viewer presented fixed SVG diagrams through an HTML and JavaScript interface and did not support model authoring or profile validation. Since the diagrams were recreated in draw.io, profile conformance was ensured manually. Moreover, it covered only one diagram type, leaving open whether the observed benefits extend to other supported types. Finally, traceability relies on manually maintained \taggedValues\ and relational \stereotypes, whose consistency cannot be guaranteed at scale without dedicated tool support.


\section{\uppercase{Conclusion and Future Work}}\label{sec:conclusion}
This paper presented the design and evaluation of SA-UML, a UML profile that enables architects to integrate \sads\ directly into the overall UML \ad, avoiding the redundancies and synchronization issues that arise when \sads\ and overall \ad\ are maintained as independent artifacts. We demonstrated its applicability through a case study comprising a structural and behavioral SA-UML model of a fictitious software system.

An evaluation with $n=27$ participants showed significantly shorter completion times in the SA-UML condition ($t(26)=2.81$, $p=0.005$, $d_z=0.54$). However, because the conditions were not information-equivalent, this result reflects the combined effect of additional security information and its integrated representation and cannot be attributed to the notation alone.
Overall, the results suggest that SA-UML is a promising approach for modeling integrated security architecture descriptions, though response-accuracy gains remain unconfirmed.

\subsection{Answering the Research Questions}

\textbf{RQ1:} We have successfully implemented a UML profile for the concepts that we proposed in~\cite{Sabau2024} using SA-UML. SA-UML was developed by systematically mapping the author's proposed concepts onto existing UML metaclasses. This required adapting some of the original relationships into a uniform \traceabilityChain, prioritizing consistent navigation between security concepts and architectural elements. \stereotypes, \taggedValues, and \constraints\ can represent \threats, \srs, \countermeasures, and their relationships, enabling traceability within the \sad. This allows us to present the security-related parts of an \ad\ (i.e., the \sad) using appropriate visualization techniques.
 
\noindent \textbf{RQ2:} The questionnaire study indicates a positive trend: participants completed the security-focused tasks significantly faster in the SA-UML condition and perceived this approach as useful.
However, no significant improvement in response scores was observed with SA-UML. The simplicity of the architectural description in the case study and learning effects may explain this, as some participants also noted.
Given the non-equivalent information content of the two diagram conditions, the findings for RQ2 should therefore be interpreted only as a preliminary trend.



\subsection{SA-UML Modeler}\label{sub:prototypes}

To fulfill our fifth objective and gain hands-on experience with SA-UML, we incorporated the evaluation's core findings into the "SA-UML Modeler", a first functional prototype UML modeling tool implemented as a lightweight HTML5 application. It loads the document portal example from section~\ref{sec:demonstration} and supports model creation, traceability analysis, and JSON import and export; as it is intended solely for exploratory purposes, it does not yet include a persistence layer. The prototype can be accessed \href{https://sa-uml-modeler.netlify.app}{online}\footnote{\url{https://sa-uml-modeler.netlify.app}}.

\subsection{Future Work}
Several directions remain open for future work. SA-UML can be extended to support additional UML diagrams, and the existing JSON-based SA-UML Modeler can be evolved into a UML-metamodel-conformant tool that automatically enforces profile constraints, with its authoring and traceability functions evaluated on larger models. The case study and evaluation could also be strengthened through a more comprehensive or industrial case study and a broader expert evaluation covering multiple architectural models in SA-UML.

\section*{Acknowledgement}
We sincerely thank all participants for their contribution to the study. Furthermore, the authors acknowledge the use of Sonnet 5.4, GPT-6 Astra, and DeepL during the preparation of this manuscript solely for language editing and improving readability. Both tools were used to refine grammar and phrasing in the paper. All results, final content, data, and scientific conclusions were developed and verified by the authors, who remain solely responsible for the integrity of the work.

\bibliographystyle{unsrt}  
\bibliography{Bibliography}

\end{document}